\documentclass[prb,superscriptaddress,reprint,amssymb,aps,floatfix,showkeys]{revtex4-1}
\usepackage{amsmath}%
\usepackage{amsfonts}%
\usepackage{amssymb}%
\usepackage{graphicx}
\usepackage{color}
\usepackage{tabularx}
\usepackage{braket}

\begin{document}
\title{Ultrasensitive microwave spectroscopy of paramagnetic impurities of sapphire crystals at millikelvin temperatures}

\author{Warrick G. Farr}
\affiliation{ARC Centre of Excellence for Engineered Quantum Systems, University of Western Australia, 35 Stirling Highway, Crawley WA 6009, Australia}

\author{Daniel L. Creedon}
\affiliation{ARC Centre of Excellence for Engineered Quantum Systems, University of Western Australia, 35 Stirling Highway, Crawley WA 6009, Australia}

\author{Maxim Goryachev}
\affiliation{ARC Centre of Excellence for Engineered Quantum Systems, University of Western Australia, 35 Stirling Highway, Crawley WA 6009, Australia}

\author{Karim Benmassai}
\affiliation{Unit\'e de Recherche en Optique et Photonique, Centre de D\'eveloppement des Technologies Avanc\'ees, S\'etif, Algeria}

\author{Michael E. Tobar}
\email{michael.tobar@uwa.edu.au}
\affiliation{ARC Centre of Excellence for Engineered Quantum Systems, University of Western Australia, 35 Stirling Highway, Crawley WA 6009, Australia}


\begin{abstract}
Progress in the emerging field of engineered quantum systems requires the development of devices that can act as quantum memories. The realisation of such devices by doping solid state cavities with paramagnetic ions imposes a trade-off between ion concentration and cavity coherence time. Here, we investigate an alternative approach involving interactions between photons and naturally occurring impurity ions in ultra-pure crystalline microwave cavities exhibiting exceptionally high quality factors.
We implement a hybrid Whispering Gallery/Electron Spin Resonance method to perform rigorous spectroscopy of an undoped single-crystal sapphire resonator over the frequency range 8--19~GHz, and at external applied DC magnetic fields up to 0.9~T. Measurements of a high purity sapphire cooled close to 100 mK reveal the presence of Fe$^{3+}$, Cr$^{3+}$, and V$^{2+}$ impurities. A host of electron transitions are measured and identified, including the two-photon classically forbidden quadrupole transition ($\Delta m_s =2$) for Fe$^{3+}$, as well as hyperfine transitions of V$^{2+}$.
\end{abstract}
\date{\today}
\maketitle


\section*{Introduction}

Single crystal sapphire ($\alpha$--Al$_2$O$_3$) is an important material used extensively in many areas of scientific research, particularly in optical, semiconductor and microwave technologies. In these fields, one of its major roles is as the gain medium in masers\cite{Maiman1960} and lasers\cite{Moulton:86}, where it is intentionally doped with a high concentration of paramagnetic impurity ions. The high concentration case has been well studied, however the role of extremely dilute, naturally occurring impurity ions contained in ultra-high purity sapphire is less well known. Such unavoidable paramagnetic impurity ions exist at a concentration of parts per billion to parts per million and have been particularly useful in a number of precision microwave experiments\cite{creed2010,5871210,locke:2737}, allowing Cryogenic Sapphire Oscillator technology to operate at its highest stability, and more recently facilitating the discovery of a number of nonlinear effects\cite{PhysRevLett.108.093902,PhysRevLett.109.143902}.

Besides application to ultra-stable frequency sources, such systems are also eminently applicable in the framework of quantum measurement, control and computation. Naturally occurring impurities in `pure' crystal resonators could lead to the realisation of microwave cavity QED (Quantum Electrodynamics) experiments in the strong coupling regime\cite{PhysRevLett.110.157001,PhysRevLett.107.060502,wirth:262508} with unprecedented coherence times. For comparison, current technology in the realm of qubit devices consists mainly of on-chip superconducting transmission line cavities with state-of-the-art coherence times on the order of several tens of microseconds. Recently, hybrid quantum systems have been proposed that involve coupling of solid state qubits, crystals intentionally doped with rare-earth ions, to microwave cavity modes in macroscopic classical resonators\cite{PhysRevB.84.060501}. Such systems achieve spin and mode coherence times of the order of seconds and microseconds respectively. The latter can be improved by using an ultra-pure single crystal sapphire operating in a Whispering Gallery (WG) mode of resonance, offering coherence times on the order of 50 ms for a typical mode with $f=11$ GHz and $Q=2\times10^9$. Despite a very low concentration of impurities, the ion-cavity photon interaction is clearly resolvable due to the extremely high quality factor of the WG modes.   
Indeed, strong interactions between microwave photons and impurity spins in cavities based on `pure' crystals do not lead to an increase in microwave losses due to artificial doping. 
In addition, such media could be used to create a quantum interface between microwave quantum processing devices and optical communication links by using coupling between microwave and optical transitions in impurity ions. All of these challenging goals require a better understanding of ion energy level transitions, and ion-photon interaction in extremely low loss, ultra-pure systems.

Unlike other spectroscopic studies of doped sapphire crystals, here we use an ultra-sensitive approach to detect and analyse naturally occurring, extremely low concentration ion impurities. 
Some preliminary observations of interactions with Fe$^{3+}$ ions in such crystals have been made in the past\cite{PhysRevB.87.094412,PhysRevB.79.174432,JAP:8482549,Xiao-Yu1987,PhysRev.123.1265}. Broadly speaking, the present work represents a systematic study of ion-photon interactions over the X and K$_\text{u}$ band of microwave frequencies (8-20 GHz). Such a complete spectroscopic study in this region is useful for all applications of sapphire in the microwave domain. 

\section{Spectroscopic Approach}

\begin{figure*}[t]
    \centering
        \includegraphics[width=1\linewidth]{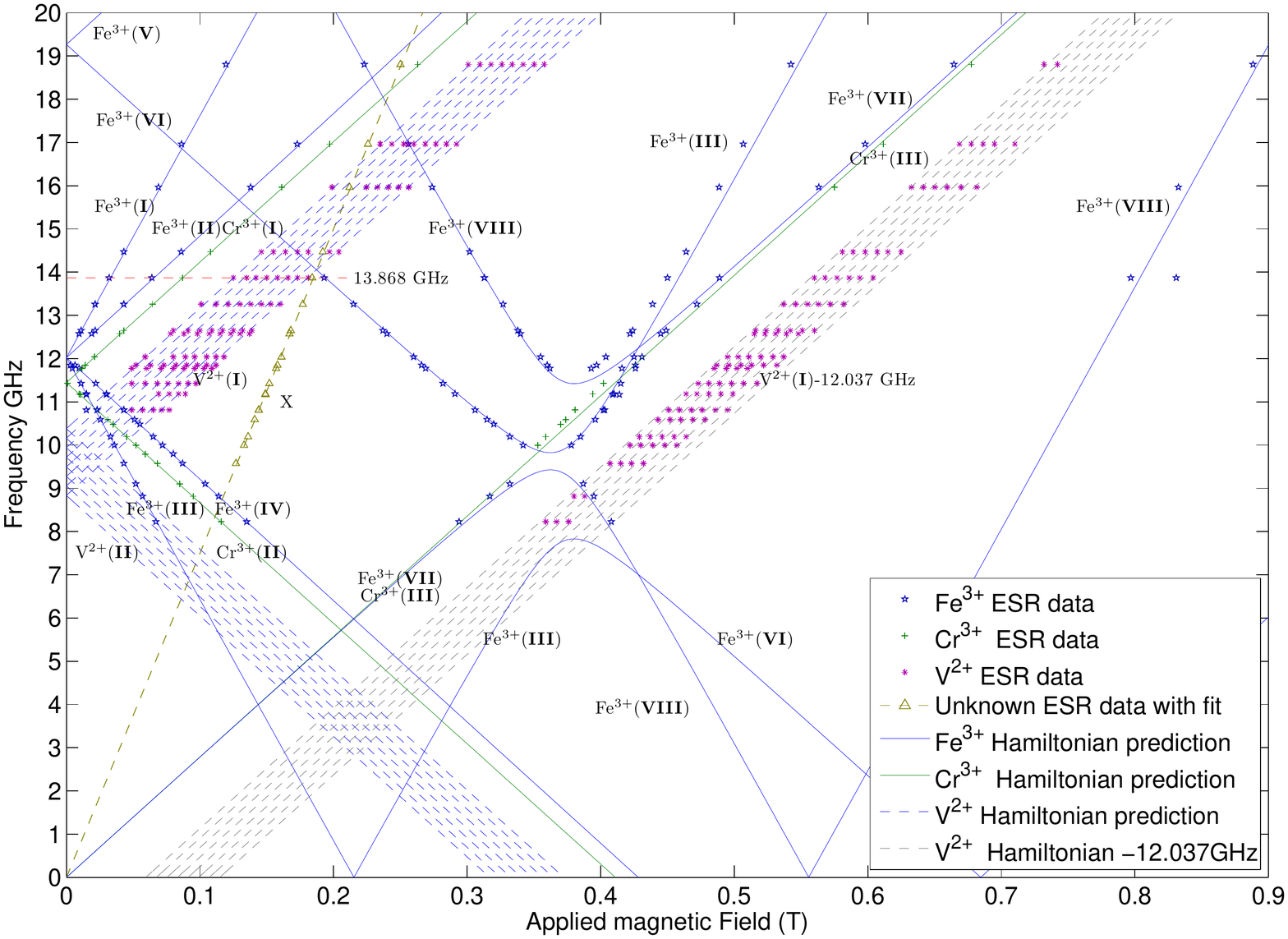}
    \caption{Interactions between numerous WG modes as a function of magnetic field. The modes are listed in table \ref{tab:interactionstrongfields2}. The admixture of $\Ket{\pm\frac{1}{2}}$ and $\Ket{\pm\frac{5}{2}}$ at high magnetic fields is neglected. }
    \label{fig:interactionstrongfields2}
\end{figure*}

In this work, we cool large cylindrical sapphire resonators close to 100 mK and use a novel technique exploiting extremely high $Q$-factor Whispering Gallery modes as a sensitive probe of Electron Spin Resonance frequencies under the influence of an external magnetic field. By sweeping an applied magnetic field and identifying avoided crossings between photonic Whispering Gallery modes and Electron Spin Resonances (ESR) of paramagnetic ions, we observe the classically forbidden quadrupole transition of Fe$^{3+}$ in sapphire, and describe the interaction of the inhomogeneously broadened Fe$^{3+}$ ESR with a number of paramagnetic spin species in the crystal. Sapphire crystals have been measured using this technique previously \cite{PhysRevB.87.094412}, however these measurements were of low sensitivity.  In the present work, we refine the technique and present spectroscopic measurements at higher sensitivity than previously achieved.  One factor in particular which led to the poor sensitivity in previous experiments was the absence of the usual copper cavity in which sapphire WG mode resonators are typically mounted to contain the stray field and ensure the highest $Q$-factor. The bore size of the superconducting magnet was too narrow to allow a cavity to enclose the crystal, which resulted in lowered WG mode $Q$-factors and thus lower sensitivity to weaker interactions. In addition, previous measurements were performed with a non-programmable magnet controller and thus the throughput and resolution of data was limited by having to continuously manually set the target field of the magnet. Finally, previous measurements focused on fields and frequencies predicted by the Fe$^{3+}$ Hamiltonian, and thus interactions in regions away from this specific Fe$^{3+}$ interaction were ignored. All these improvements lead to more systematic study of dilute impurities and their interaction with the field. 


\subsection{Measurement Methodology}

To characterize the various species of paramagnetic ions in sapphire, it is possible to record the frequency and magnetic field response of a range of WG modes supported in the resonator. As a DC magnetic field is applied parallel to the $c$-axis of the crystal and swept in magnitude, the sapphire cavity transmission coefficient $S_{21}$ is measured for a spectrum of WG modes. The modes, with $Q$-factors commonly in the range of $10^8 - 10^9$ in sapphire, act as extremely narrow band pass filters sharply defined in frequency when far-detuned from an ESR transition. When the ESR is tuned close to the WG mode resonance by the DC external magnetic field, the $S_{21}$ trace shows a shift in the WG mode frequency and a degradation of its Q factor. Fits were made to the $S_{21}$ data to find the lowest value of amplitude, corresponding to the field at which the ESR transition is tuned on resonance with the WG mode.

The modes were measured with an applied external DC magnetic field with a strength up to 0.9~T. Whispering Gallery modes act as extremely sensitive probes of the presence of impurity spins 
due to considerable photon-spin interaction. The strength of this interaction depends on the mode polarisation (WGH, WGE or hybrid), mode volume, mode quality factor and ion concentration. Generally higher quality factors are preferable because they increase photon-spin interaction probability as photon lifetime in the cavity increases. 
 By measuring a spectrum of WG modes across a broad frequency range, it is then possible to determine energy levels of corresponding impurity ions, Lande $g$-factors, photon-spin coupling, impurity concentration, ESR linewidths, etc. 

\subsection{Experiment Details}

\begin{figure}
  \includegraphics[width=1\linewidth]{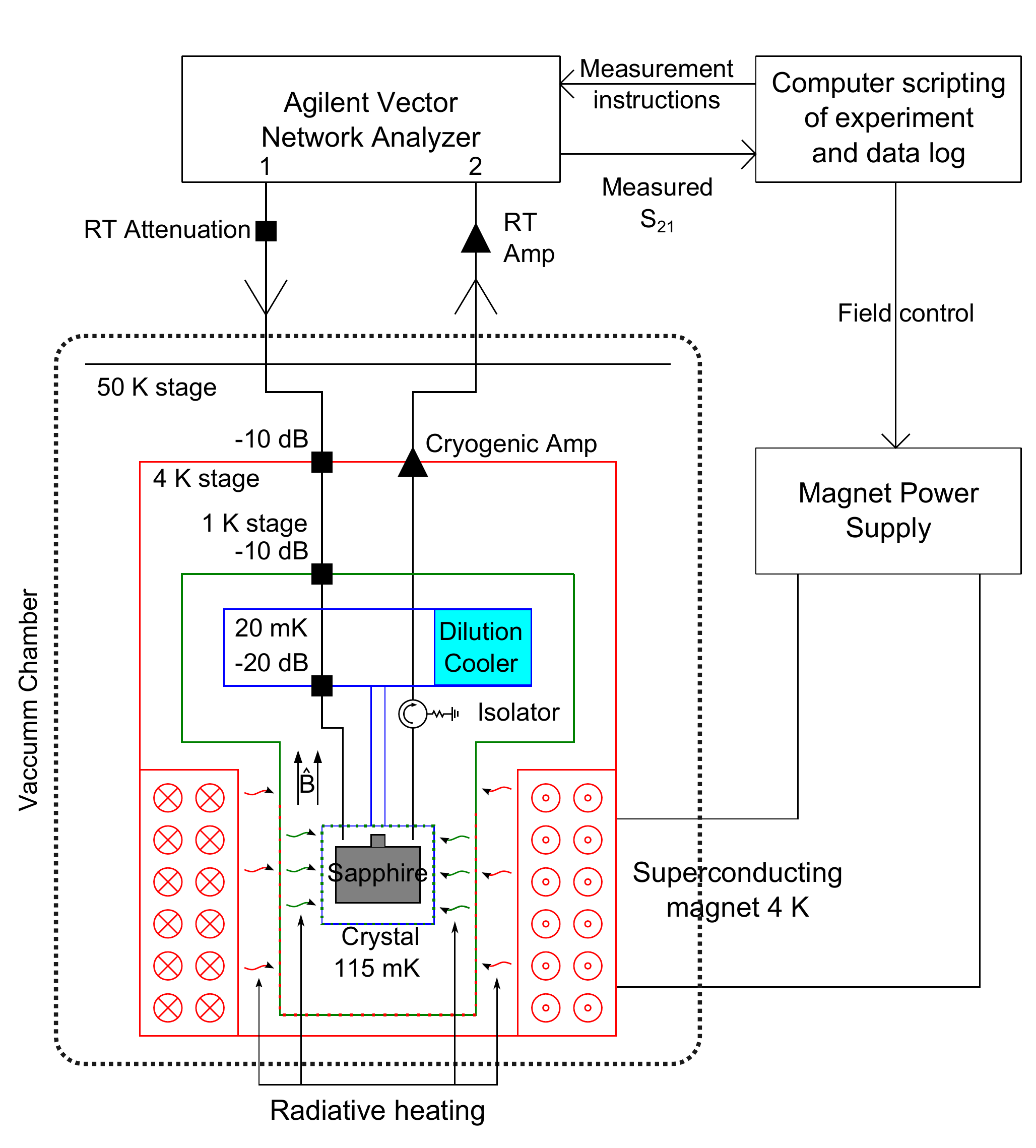}\\
  \caption{The experimental apparatus. The sapphire resonator was mounted within an oxygen free copper cavity and measured in transmission with between -90~dBm and -30~dBm incident power from a network analzer. The crystal is cooled to approximatley \mbox{115 mK} by the dilution refrigerator. The super conducting magnet is controlled externally by a computer.}\label{ea2}
\end{figure}

The crystal under study was a highest-purity sapphire from GT Crystal Systems (USA) grown using the heat exchange method (HEMEX) and machined into a cylinder 50 mm diameter $\times$ 30 mm height. The concentration of Fe$^{3+}$ impurity ions in the crystal has been measured previously to be approximately 100 parts per billion\cite{0022-3727-33-6-301}.  Such a concentration was achieved through an annealing process in air which caused mass conversion of Fe$^{2+}$ to Fe$^{3+}$. The crystal is cut such that the $c$-axis of the crystal structure is oriented is parallel to the cylindrical $z$-axis of the resonator. To attain the highest quality factor whispering gallery modes at low temperature, the crystal is mounted in a cavity manufactured from oxygen-free high thermal conductivity copper to avoid hot-spots at oxygen impurities. Energy is coupled to the crystal through two straight antennae orientated 180$^{\circ}$ from each other, entering through the top of the cavity and parallel to the $z$ axis of the crystal. The crystal was cooled using a cryogen-free dilution refrigerator (DR), with a cooling power of 1.5 W at the second pulse tube stage (4 K), and $\sim$500 $\mu$W at 100 mK (see Fig.~\ref{ea2}). A superconducting magnet was attached to the 4 K stage of the DR, with the crystal mounted within a 1~K radiation shield which sits within the bore of the magnet. Due to the small diameter of the magnet bore, the distance between the outer wall of the cavity and the inner wall of the 1~K shield inside the bore is approximatley 3~mm, permitting weak thermalization between the 1K~shield and the crystal. Although the crystal is attached to the mixing chamber plate of the fridge ($\sim$20 mK), after thermalization the crystal reaches an equilibrium temperature of $\sim$115~mK. The corresponding temperature sensor is directly attached to a copper block holding the crystal.

A number of WG modes were measured in the resonator using a vector network analyzer at an excitation power between -90 and -30~dBm incident on the input probe antenna. At higher powers, the temperature of the crystal increases and nonlinear behaviour begins to occur in the crystal. The network analyzer in these measurements had an upper frequency limit of 20~GHz. The cables inside the DR were connected to each stage via a feedthrough and DC block to allow thermalization of both the inner and outer conductor, and to reduce temperature gradients in the system. In addition, 10~dB attenuators were included in the line at the 4~K stage and 1~K stage of the DR, and a 20~dB attenuator was attached at the mixing chamber stage. Cold attenuation ensures a sufficiently low power signal is incident on the crystal without the addition of thermal noise from room temperature attenuation. 
The signal transmitted through the WG cavity passes through a low noise cryogenic amplifier ($4$K stage) separated from the cavity by an isolator ($20$mK stage) preventing the back action noise. Another amplification cascade is used at the room temperature to ensure a better signal-to-noise ratio.   

MATLAB software was developed to automatically control the field strength of the magnet, and the sweep parameters and data acquisition of the network analyzer. The sweep rate of the magnetic field was set slowly enough that temperature of the copper cavity enclosing the sapphire was not significantly changed by induced eddy currents. A ramping rate of $\tfrac{dB}{dt}=0.005$~T/min was found to successfully ensure that all measurements were made at thermal equilibrium. A schematic of the experimental apparatus is shown in Fig.~\ref{ea2}. The frequency stability of the measurement system is maintained by a hydrogen maser to avoid all long term fluctuations during the experiment.

\section{Theoretical Description of Impurity Ions in Sapphire}

\begin{figure}
	\centering
		\includegraphics[width=1.00\linewidth]{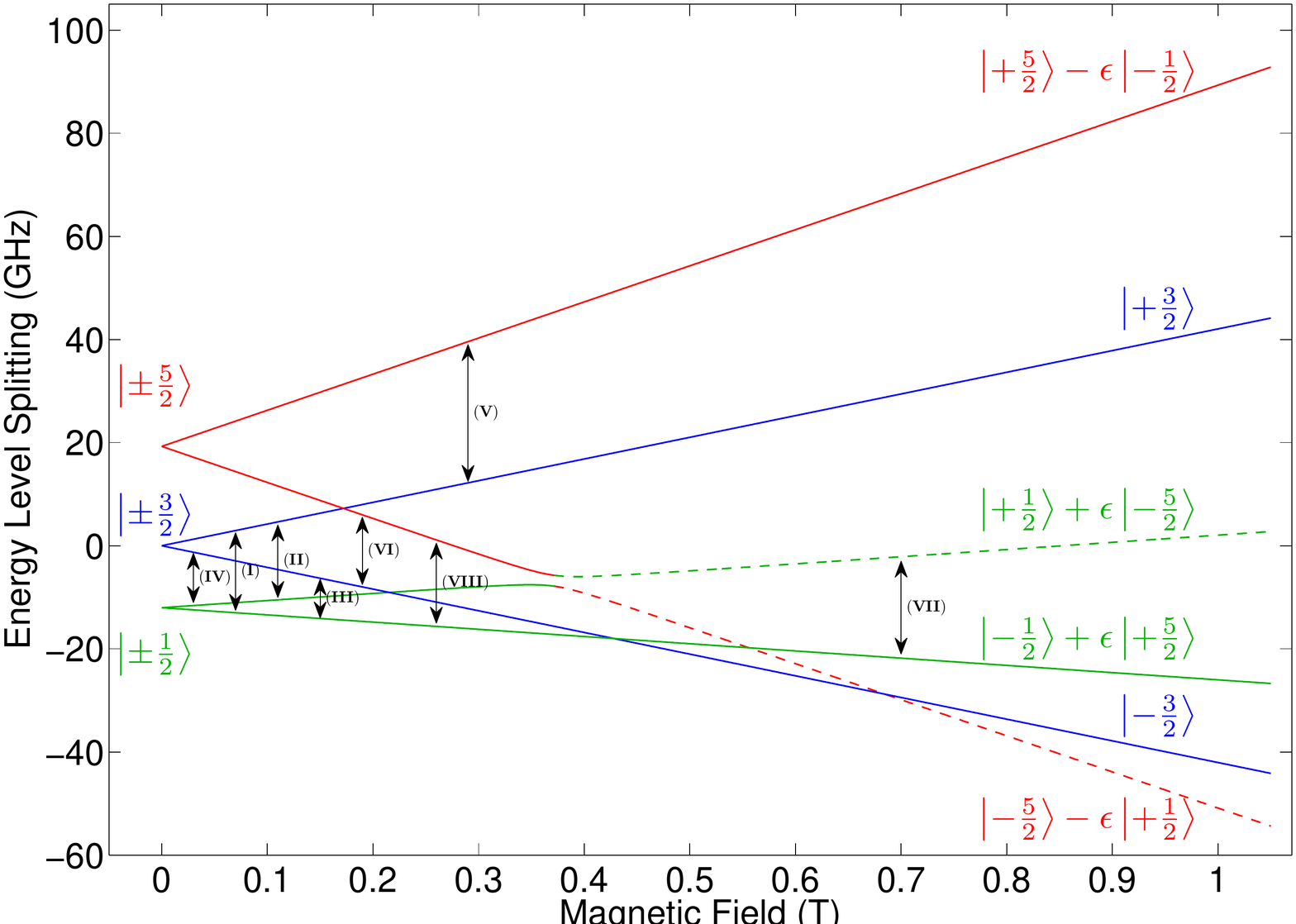}
	\caption{Energy level splitting of the Fe$^{3+}$ ion with applied magnetic field. The calculation demonstrates an admixture of spin states $\Ket{\pm1/2}$ and $\Ket{\pm5/2}$ for fields above.}
	\label{fig:hamiltonianfe3p}
\end{figure}

The most naturally abundant isotope of Iron is $^{56}$Fe, with a nuclear spin $I=0$. The fundamental energy levels of the Fe$^{3+}$ ion in sapphire are defined by the zero-field splitting parameter $^6$S in the spin Hamiltonian. Fe$^{3+}$ has been extensively studied in sapphire\cite{Bogle.0370-1328-73-3-425,symmons.2}Three levels exist at zero applied DC magnetic field, with dependence on a DC magnetic field described by the following Hamiltonian:


\begin{multline}
\mathcal{H}_\text{Fe}=\beta\hat{\textbf{H}}\textbf{g}\hat{\textbf{S}}+D\left[\hat{\textbf{S}}_{z}^{2}-\frac{1}{3} S\left(S+1\right)\right]\\
+\tfrac{1}{6}a\left[\hat{\textbf{S}}_{\xi}^{4}+\hat{\textbf{S}}_{\eta}^{4}+\hat{\textbf{S}}_{\zeta}^{4}\right]-\tfrac{1}{5}\left[S\left(S+1\right)\left(3S^2+3S-1\right)\right]\\
+\tfrac{1}{180}F\left[35\textbf{S}_{z}^{4}-30S(S + 1)\hat{\textbf{S}}_{z}^{2} + 25\hat{\textbf{S}}_{z}^{2} - 6S(S + 1)\right.\\
\left.+3S^2(S+1)^2\right],
\label{eq1}
\end{multline}
where $a$ is the (fourth order) spin Hamiltonian parameter, $g$ is the g factor tensor, $\beta$ is the Bohr magneton, $\textbf{H}$ is the vector magnetic field strength, $\hat{\textbf{S}}$ is the electronic spin operator where $\hat{\textbf{S}}_i$ are the components of the electronic spin operator, $D$ is the (second order) zero field splitting (ZFS) parameter, $F$ is the (fourth order) ZFS parameter. The first term in Hamiltonian (\ref{eq1}) is a typical Zeeman splitting term. The second one is known as single-ion anisotropy term describing interaction of the ion spin angular momentum with the lattice field via the orbital angular momentum. The corresponding parameters $D$ characterises ZFS caused by the crystal field. The other terms represent higher order interactions between the crystal lattice and the Fe$^{3+}$.
The described Hamiltonian parameters from the previous studies are summarized in Table \ref{hamfe}.
An analytical expression for the energy levels of Fe$^{3+}$ in sapphire when a magnetic field is applied parallel to the $c$-axis of the crystal can also be derived:
\begin{align}
W_{\Ket{\pm\frac{1}{2}}}=	&  \pm g_{||}\beta{H}_{z}-\frac{3}{2} (a-F)+ D \notag\\
																&-\frac{1}{6}\sqrt{\left[\mp9 g_{||}\beta{H}_{z}+(a-F)+18 D\right]^2+80a^2},\notag\\
W_{\Ket{\pm\frac{3}{2}}}=	& \frac{\pm3}{2} g_{||}\beta{H}_{z},  \notag\\
W_{\Ket{\pm\frac{5}{2}}}=	& \pm g_{||}\beta{H}_{z}-\frac{3}{2} (a-F)+ D \notag\\
																&+\frac{1}{6}\sqrt{\left[\pm9 g_{||}\beta{H}_{z}+(a-F)+18 D\right]^2+80a^2},
\end{align}
where the energy levels are described here for the case of low magnetic field. At 0.38~T, the $\Ket{\pm\frac{5}{2}}$ and $\Ket{\mp\frac{1}{2}}$ states mix, as shown in Fig. \ref{fig:hamiltonianfe3p}.

\begin{figure}
	\centering
		\includegraphics[width=1.00\linewidth]{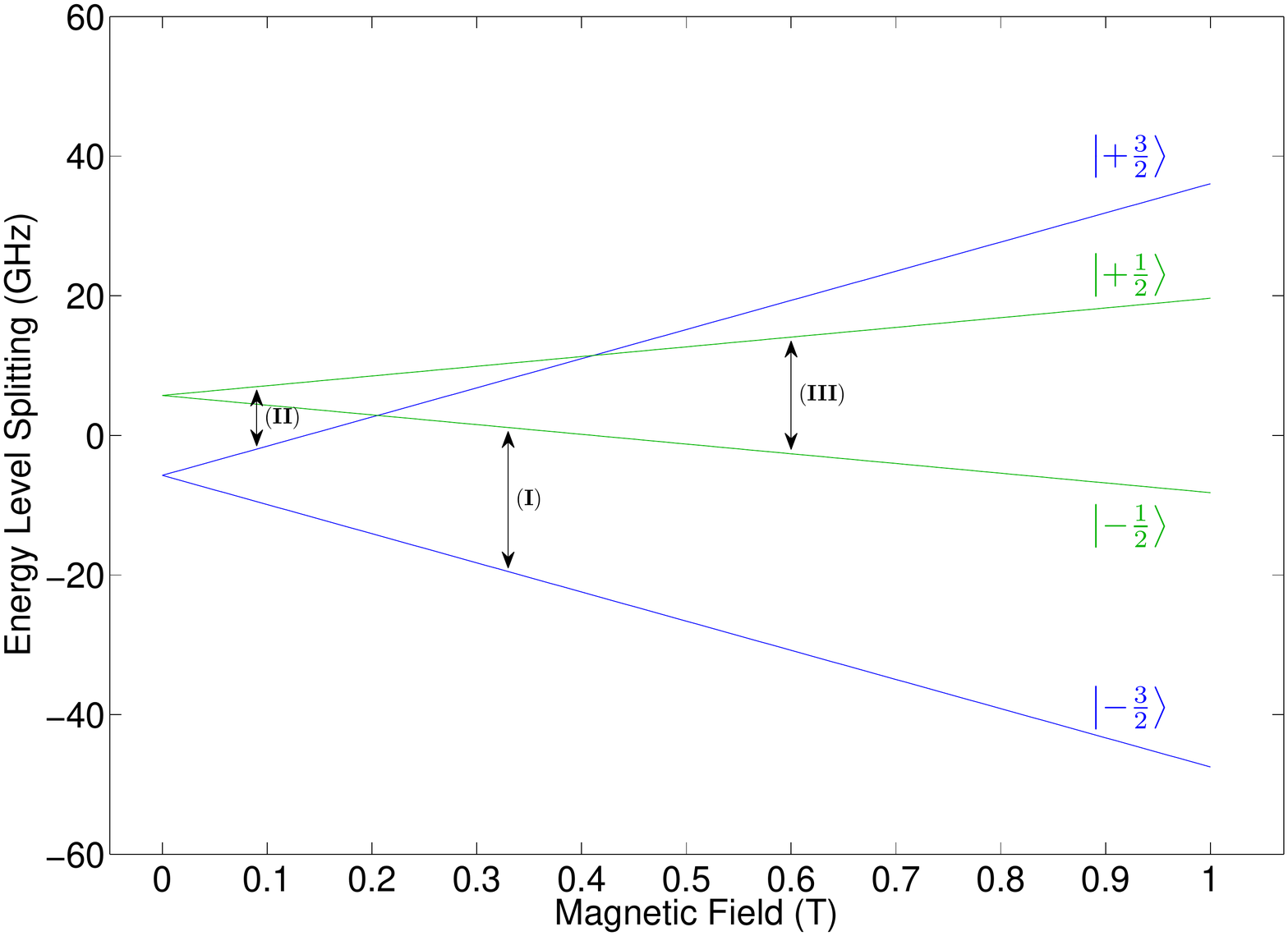}
	\caption{Energy level splitting of the Cr$^{3+}$ ion. The ZFS parameter $D$ is negative which causes the higher spin to have lower energy at 0 field. }
	\label{fig:hamiltoniancr3p}
\end{figure}

In addition to iron, sapphire also contains chromium impurities of the type Cr$^{3+}$ at a concentration of parts per billion. Whilst chromium does have an isotope with nuclear spin ($^{53}$Cr, with $I = \tfrac{3}{2}$), the $^{52}$Cr isotope has a much higher natural abundance (83.8\%) and possesses zero nuclear spin. This makes it possible to ignore nuclear spin terms in its Hamiltonian which reduces to the Zeeman and zeroth order single-ion anisotropy term described for the case of Fe$^{3+}$:
\begin{align}
\mathcal{H}_\text{Cr}=&\beta \hat{\textbf{H}}\textbf{g}\hat{\textbf{S}} +D[\hat{\textbf{S}}_z^2-\tfrac{1}{3}S(S+1)].
\end{align}
The corresponding energy levels are demonstrated in Fig.~\ref{fig:hamiltoniancr3p}. 

 Another natural ion impurity of sapphire crystals, Vanadium, has only one stable isotope $^{51}$V with abundance (99.75\%), which is isoelectronic\cite{Stedman1969} with Cr$^{3+}$ and has nuclear spin $I = \tfrac{7}{2}$. Vanadium impurities exist in sapphire as V$^{2+}$, normally due to conversion from the trivalent state resulting from x-ray or gamma irradiation\cite{Lambe1960}. However, their presence in the crystal under study is more likely to be a result of the annealing process as the crystal has never undergone radiation treatment. The nuclear spin of the vanadium gives rise to eight hyperfine lines\cite{Abragam1951} under the application of a magnetic field. 
 Vanadium ions can be described by the spin Hamiltonian\cite{PhysRev.123.1265}:
\begin{align}
\mathcal{H}=&\beta \hat{\textbf{H}}\textbf{g}\hat{\textbf{S}} +D[\hat{\textbf{S}}_z^2-\tfrac{1}{3}S(S+1)]
+A\hat{\textbf{S}}_z\hat{\textbf{I}}_z\notag\\&+B[\hat{\textbf{S}}_x\hat{\textbf{I}}_x+\hat{\textbf{S}}_y\hat{\textbf{I}}_y]-\gamma_n\beta_n\textbf{H}\hat{\textbf{I}}\notag\\=
&+Q'[\hat{\textbf{I}}_z^2-\tfrac{1}{3}I(I+1)],
\end{align}
where $\hat{\textbf{I}}$ is the nuclear spin operator, $\gamma_n$ and $\beta_n$ are the $g$ factor and Borr magnetron for the nucleus spin, the $A$ and $B$ are electron-nucleous coupling constants along the $z$ and two transverse axis, $Q'$ a constant characterising the single-ion anisotropic for the nucleus.


The energy levels to first order when the field is parallel to the $c$-axis is given by the following expression:
\begin{align}
W_{\Ket{{S}_z,{I}_z}}=& g_{||}\beta {H}_{z}{S}_z+D[{S}_z^2-\tfrac{1}{3}S(S+1)]+A{S}_z{I}_z\notag\\
&-\gamma_n\beta_n {H}_{z}{I}_z+Q'[{I}_z^2-\tfrac{1}{3}I(I+1)].
\end{align}`
This first order approximation is sufficient for the measurements in the studied magnetic field range which does not include the crossing point of the energy levels.

%


\begin{figure}
	\centering
		\includegraphics[width=1.00\linewidth]{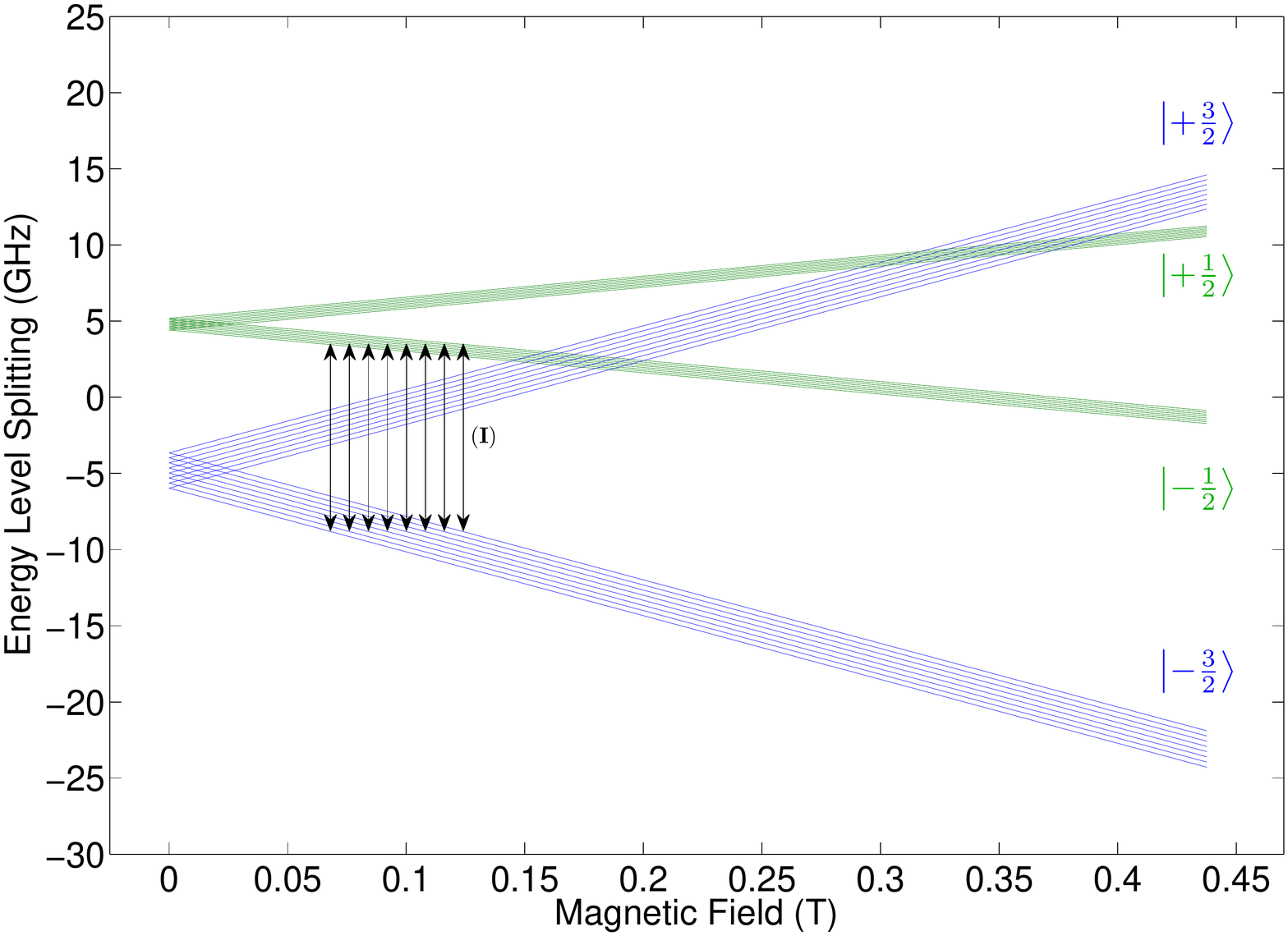}
	\caption{The V$^{2+}$ impurity ion demonstrates hyperfine splitting into eight lines.}
	\label{fig:hamiltonianV3p}
\end{figure}
\begin{table}
\caption{Hamiltonian parameters of ion impurities as measured previously at approximately 4~K.}

\begin{tabular}{|rrrr|}
\hline
\hline
Ion&Parameter&Kornienko et al\cite{Kornienko.749982}&Symmons et al\cite{symmons.2}\\
\hline
$^{56}$Fe$^{3+}$&S&5/2&\\
&I&0&\\
&$D$ (MHz)&$5146.5\pm2$&$5157.6\pm3$\\
&$\|a\|$ (MHz)&$709.6\pm 4$&$688.2\pm12$\\
&$a-F$ (MHz)&$1015.3\pm3$&$ 1025\pm3$\\
&$g_{||}$&$2.0029\pm0.0007$&$2.0026\pm0.0005$
\\\hline
Ion&Parameter&Laurance\cite{PhysRev1963_132_1029_1036}&\\
\hline
$^{52}$Cr$^{3+}$&S&3/2&\\
&I&0&\\
&$D$(MHz)&-5723.5$\pm3$&\\
&$g_{||}$&1.984&
\\\hline
Ion&Parameter&Laurance\cite{PhysRev1963_132_1029_1036}&\\
\hline

$^{51}$V$^{2+}$&S&3/2&\\
&I&7/2&\\
&$D$ (MHz)&$-4803.6\pm1.0$ &\\

&$A$ (MHz)&$-220.615\pm0.025$&\\
&$B$ (MHz)&$-222.890\pm0.100$&\\
&$g_{||}$&1.991&\\\hline

\end{tabular}

\label{hamfe}

\end{table}

\section{Main Results}
Using the technique discussed above, the sapphire crystal has been tested using various modes for the range of applied DC magnetic fields. 

\subsection{WGH$_{20,0,0}$ mode}

\begin{figure}
  \includegraphics[width=1\linewidth]{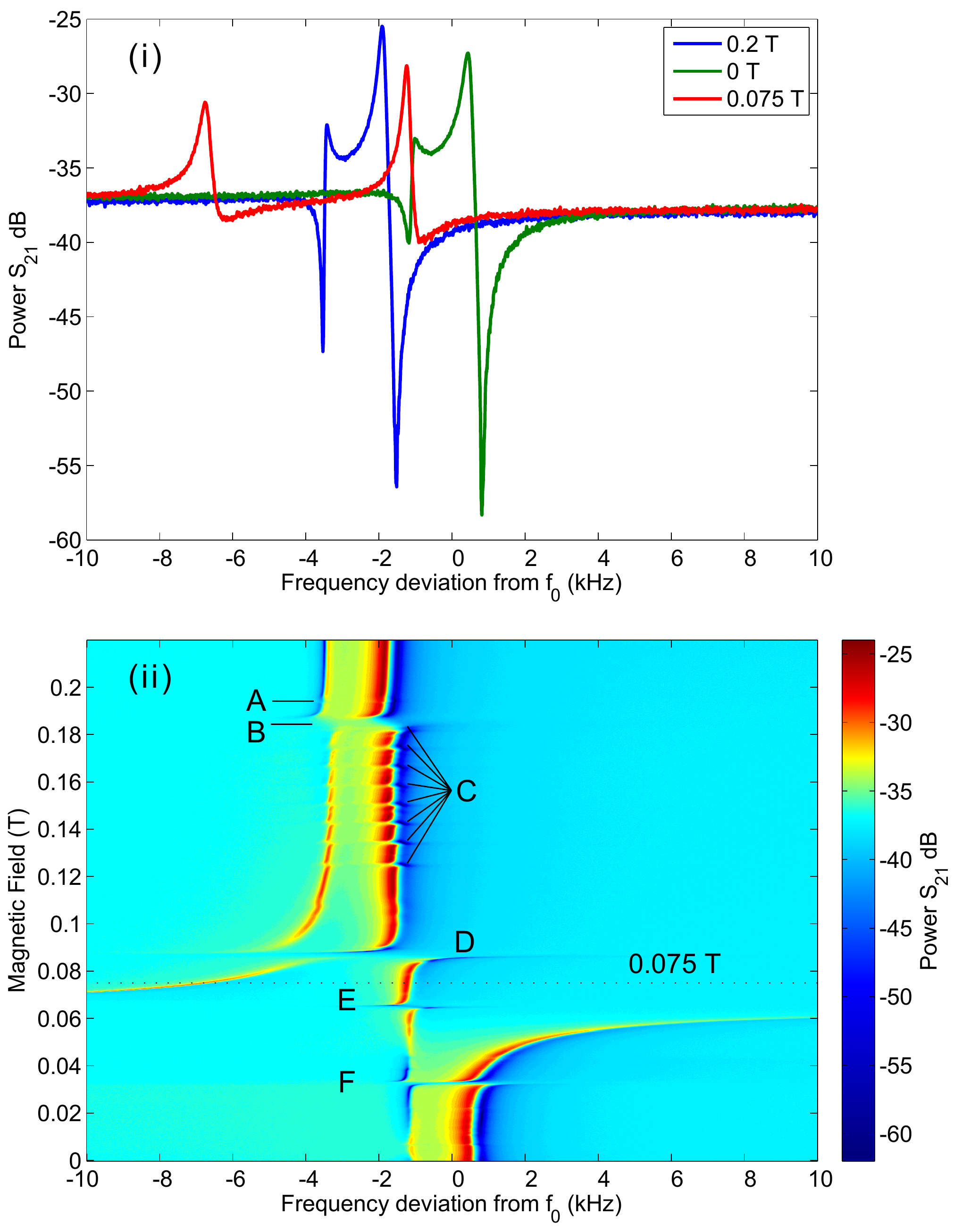}\\
  \caption{ (i): cavity transmission at various magnetic fields; (ii): Magnetic field dependence of the WGH$_{20,0,0}$ mode doublet . A number of clear interactions can be seen between the mode and electron spin resonances of paramagnetic impurity ions in the crystal. The six interactions shown (labeled A to F) are identified in Table \ref{tablemodeat13}.} 
  \label{whigmo}
	
\end{figure}
\begin{table}[t]
\caption{Spin-photon interactions of the WGH$_{20,0,0}$ from Figure \ref{whigmo}. Although the interaction at F is classically forbidden ($\Delta m_s=2$), it has a non-zero probability of occurring \cite{jrnbsLewiner1969}.}
\centering
\begin{tabularx}{\columnwidth}{ccXX}
\hline
\hline
& Ion&Transition&Description\\
\hline
A& Fe$^{3+}$ &$\Ket{+\tfrac{5}{2}}\rightarrow\Ket{+\frac{3}{2}}$& Dipole\\
B& Unknown &$\Ket{\pm\tfrac{x}{2}}\rightarrow\Ket{\mp\frac{x}{2}}$& Spin flip\\
C& V$^{2+}$&$\Ket{-\tfrac{3}{2}}\rightarrow\Ket{-\frac{1}{2}}$& Hyperfine dipole \\
D& Cr$^{3+}$&$\Ket{-\tfrac{3}{2}}\rightarrow\Ket{-\frac{1}{2}}$& Dipole\\
E& Fe$^{3+}$&$\Ket{+\tfrac{3}{2}}\rightarrow\Ket{+\frac{1}{2}}$& Dipole \\
F& Fe$^{3+}$&$\Ket{+\tfrac{3}{2}}\rightarrow\Ket{-\frac{1}{2}}$& Quadrupole \\
\hline
\end{tabularx}
\label{tablemodeat13}
\end{table}

The 20$^{\text{th}}$ azimuthal order WG mode, labeled as WGH$_{20,0,0}$, is given here as an example of a known high-$Q$ WG mode strongly interacting with different ion ensembles occurring in the crystal lattice. A two-fold degeneracy in the azimuthal ($\phi$) component of the WG mode solution, having a basis of the form $e^{-i\phi}$ and $e^{i\phi}$, is lifted by various symmetry breaking effects in the cylindrical cavity. These effects result in the mode existing as a doublet. The WGH family of modes is characterised by having the maximum magnetic component along the radial axis of the crystal cylinder. Thus, since the DC magnetic field is always axial in this cylinder, microwave and DC magnetic field components are normal to each other, making this family the most sensitive to ion spins.  \\

As discussed in the previous section, the WGH$_{20,0,0}$ doublet mode at 13.868,226,886~GHz is measured in transmission as a function of magnetic field.  For the crystal under test, it is known that Fe$^{3+}$ is one of the most abundant impurity ions present, which has resulted in the observation of a gyrotropic effect previously\cite{PhysRevB.79.174432}. Figure \ref{whigmo} shows the results of the measurement, in which is is possible to see a very strong interaction at 0.06~T corresponding to the (dipole) transition between the $\Ket{+\frac{3}{2}}$ and $\Ket{+\frac{1}{2}}$ spin states of the Fe$^{3+}$ ion. In addition, at 0.03~T the classically forbidden quadrupole (two-photon) transition for Fe$^{3+}$ is observed. With increasing magnetic field, an interaction corresponding to the $\Ket{-\frac{3}{2}}\rightarrow \Ket{-\frac{1}{2}}$ transition of Cr$^{3+}$ - an impurity spin with much smaller concentration than Fe$^{3+}$. The set of 8 closely spaced interactions are due to hyperfine splitting due to the presence of a V$^{2+}$ spin ensemble. Finally, two remaining interactions are observed which correspond to the $\Ket{+\frac{5}{2}} \rightarrow \Ket{+\frac{3}{2}}$ and $\Ket{+\frac{3}{2}} \rightarrow \Ket{-\frac{3}{2}}$ transitions of Fe$^{3+}$.\\

The dipole transition $\Ket{+\frac{1}{2}} \rightarrow \Ket{+\frac{3}{2}}$ of Fe$^{3+}$ ion shown in Fig.~\ref{whigmo}, the WG mode doublet approaches asymptotically the ESR transition line. This is a typical atom-field interaction picture, where the cavity mode is more photon-like where it is close to zero frequency deviation and more atom-like far from it. Although, finite spin-lattice coupling introduces dissipation to the ESR considerably broadening its spectrum. Thus, for Fe$^{3+}$ ions in sapphire, the width is typically 27 MHz\cite{PhysRevB.87.094412}, and it is set by crystallographic nature of the host material coupling to the ion. The spin lattice interaction appears to be stronger than the cavity photon-spin interaction preventing observation of the spin-photon strong coupling\cite{ritsch,PhysRevLett.110.157001}.

Figure \ref{whigmo} also shows a demonstration of the gyrotropic effect, different AC susceptibility for different peaks of a doublet WG mode. Each peak of the doublet represents one of two counterpropagating elliptically polarized waves of the same WG mode. These waves are coupled through imperfections in the sapphire, back scatterers, anisotropy, non-ideal geometry, etc \cite{PhysRevB.79.174432}. Each of the two modes therefore carries microwave photons of opposite spin angular momentum quantum number, or helicity, which interact with ESR transitions differently due to requirements imposed by the spin angular momentum conservation law.

\subsection{High Field Spectroscopy}

The proposed method of spectroscopy was performed on the modes of interest up to applied magnetic field strengths of 0.9~T. Such high field levels make it possible to determine the properties of the various ensembles of spin in the lattice, including the zero field splitting and $g$-factors. The results obtained for these parameters using the WGM method are shown in Fig. \ref{fig:interactionstrongfields2}, and are summarized in Table \ref{table2}.
The gradient of the frequency dependence on magnetic field of an ESR transition is fit to $\frac{df}{dB}\approx\frac{g\mu_B}{\hbar}$.

Using the WGM method, it is possible to map out the path of various ESR transitions from zero field up to any arbitrary value of magnetic field within the available frequency range. The results reveal very low abundance trace elements present in the sapphire, with estimations of the $g$-factors of these ions.  A number of ESR transitions are observed to have a zero field splitting of less than 1 GHz, making such transitions difficult to distinguish from each other except by $g$-factor measurements at higher magnetic fields. Once the modes have been identified with this method, the measurements can then be used to determine energy level transitions as predicted by the electron spin Hamiltonians.

The spectroscopy results revealed the previously unknown strong transition labeled by $X$ in Fig.~\ref{fig:interactionstrongfields2} and Table~\ref{tab:interactionstrongfields2}. The transition has unusual $g$-factor of about $5.4$ which is considerably distinct from an integer multiple of 2. This transition cannot be attributed to three photon transition $\Ket{-\frac{3}{2}}\rightarrow\Ket{+\frac{3}{2}}$ due to its low probability and significant difference from the expected $g$-factor of 2. Indeed, the corresponding ion has to be quite abandoned in the crystal and have an anomously large single-photon transition $g$-factor. Analysis of the literature on ions in sapphire crystals reveals no candidate with such characteristics. Although significant difference of the $g$-factor from an integer multiple of $2$ suggests a rare-earth nature of such impurity.

\begin{table*}
\begin{tabular}{|l||l|l|l|l|l|l|l|}
  \hline
  Ion   & Spin transition &Line&I$_{z}$& $\Delta$S$_{z}$ &$f(0)$ (GHz) & $df/dB$ &$g$-factor \\\hline
    Fe$^{3+}$   & \tiny{$\Ket{-\frac{1}{2}}\rightarrow\Ket{+\tfrac{3}{2}}$}
 &\textbf{I}&& 2 & 12.031 & 56.995      & 4.0722 \\
                & \tiny{$\Ket{+\frac{1}{2}}\rightarrow\Ket{+\frac{3}{2}}$}
 &\textbf{II}&& 1 & 12.038 & 28.392     & 2.0285 \\\cline{2-6}
                & \tiny{$\Ket{-\frac{1}{2}}\rightarrow\Ket{-\frac{3}{2}}$}
 &\textbf{III}&& 1 & 12.032 & -28.207     & 2.0153 \\
                & \tiny{$\Ket{+\frac{1}{2}}\rightarrow{\Ket{-\frac{3}{2}}}$}
 &\textbf{IV}&& 2 & 12.047 & -56.710     & 4.0518 \\\cline{2-6}
           
                & \tiny{$\Ket{+\frac{3}{2}}\rightarrow\Ket{+\tfrac{5}{2}}$}
 &\textbf{V}&& 1 &   &       &  \\
& \tiny{$\Ket{-\frac{3}{2}}\rightarrow\Ket{-\tfrac{5}{2}}$}
 &\textbf{VI}&& 1 & 18.866	 & -26.224      & 1.871 *\\\cline{2-6}
                
                & \tiny{$\Ket{-\frac{1}{2}}\rightarrow\Ket{+\tfrac{1}{2}}$}
 &\textbf{VII}&&1  &	 &       & \\
		& \tiny{$\Ket{-\frac{1}{2}}\rightarrow\Ket{-\tfrac{5}{2}}$}
 &\textbf{VIII}&& 2 & 31.911 & -57.301      &4.088 *\\\hline
		Cr$^{3+}$   & \tiny{$\Ket{\tfrac{-1}{2}}\rightarrow\Ket{-\frac{3}{2}}$}
 &\textbf{I}&& 1 & 11.454 & 27.889      & 1.9926 \\
                & \tiny{$\Ket{+\tfrac{1}{2}}\rightarrow\Ket{+\frac{3}{2}}$}
 &\textbf{II}&& 1 & 11.451 & -27.774      & 1.9844 \\
                & \tiny{$\Ket{-\tfrac{1}{2}}\rightarrow\Ket{+\frac{1}{2}}$}
 &\textbf{III}&& 1 &  &        &   \\\hline
    V$^{2+}$   & \tiny{$\Ket{-\frac{1}{2}}\rightarrow\Ket{-\frac{3}{2}}$}
								&\textbf{I}&$\Ket{-\frac{7}{2}}$& 1 & 8.677 & 28.401     & 2.0292 \\
                &&   &$\Ket{-\frac{5}{2}}$& 1 & 8.834 & 28.615     & 2.0445 \\\cline{3-6}
                &&   &$\Ket{-\frac{3}{2}}$& 1 & 9.017 & 28.765      & 2.0552 \\
               & &   &$\Ket{-\frac{1}{2}}$& 1 & 9.248 & 28.796      & 2.0574 \\\cline{3-6}
                &&  & $\Ket{\frac{1}{2}}$& 1 & 9.489 & 28.719      & 2.0519 \\
               & &   &$\Ket{\frac{3}{2}}$& 1 & 9.779 & 28.488      & 2.0354 \\\cline{3-6}
               & &   &$\Ket{\frac{5}{2}}$& 1 & 10.075 & 28.229     & 2.0169 \\
               & &   &$\Ket{\frac{7}{2}}$& 1 & 10.403 &  27.914    & 1.9944 \\\hline
X$^{+}$     &
 &&&  & -0.065 & 75.540      & 5.3972 \\\hline

\end{tabular}
\caption{Ion ensembles and corresponding g-factors, with numbers obtained from the high field spectroscopy fits. Fits marked with an asterisk are made on limited range of data points that limits the corresponding accuracy. The `Line' values identified with roman numerals correspond to the transitions shown in Fig. \ref{fig:interactionstrongfields2}.
}
\label{tab:interactionstrongfields2}
\label{table2}
\end{table*}

\begin{figure}
	\begin{centering}

  \includegraphics[width=1\linewidth]{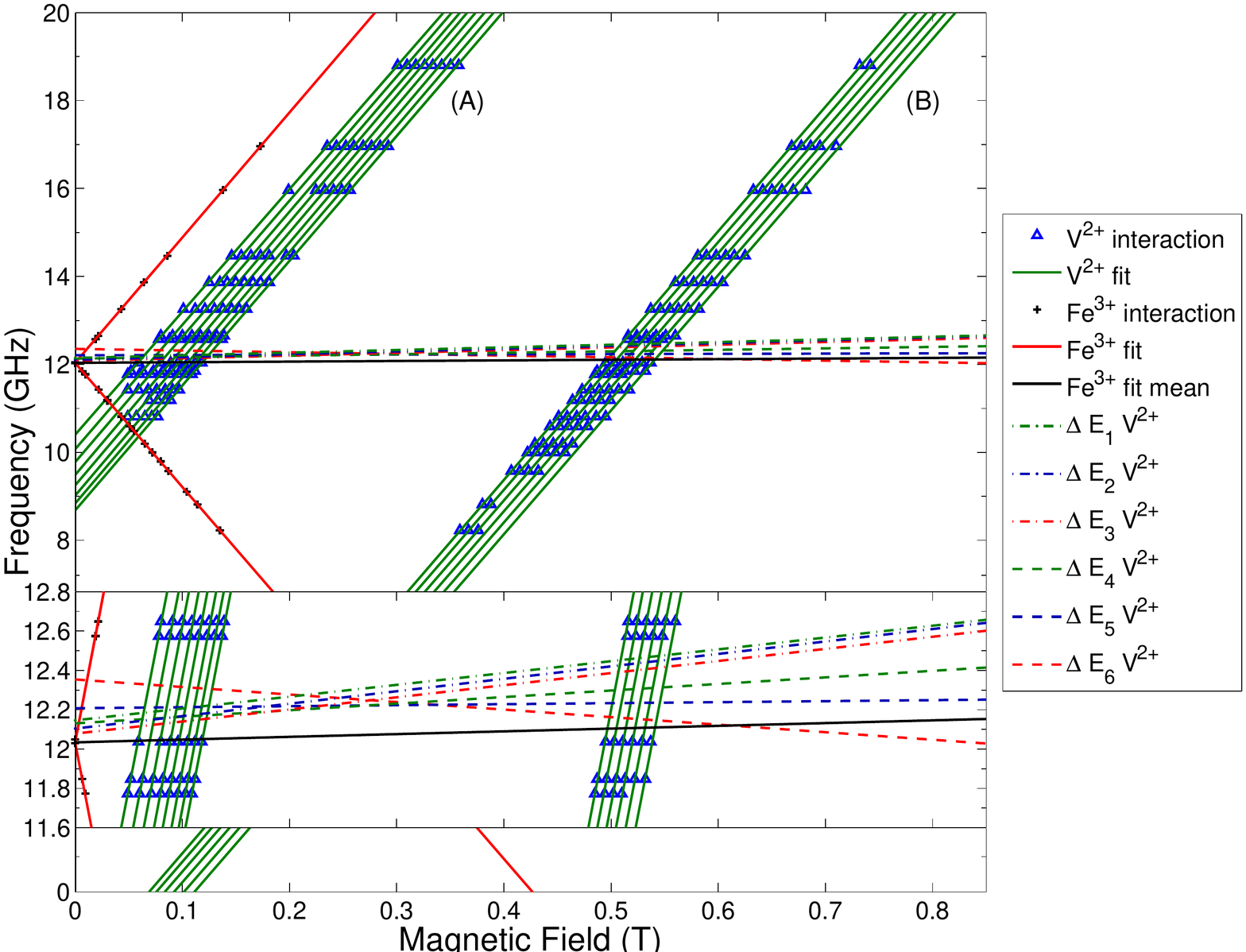}
  
  \caption{Interactions of the V$^{2+}$ ions (green lines) and Fe$^{3+}$ ions (red lines). Dashed lines demonstrate energy difference $\Delta E_{n}$ between corresponding lines $n$ from group (A) and group (B). The difference coincides with the ZFS of Fe$^{3+}$ ions.}\label{interactionstrongfields2_vanadiumspinspin2}
\end{centering}

\end{figure}

\subsection{Interaction Between Ion Ensembles}

The V$^{2+}$ ion has a nuclear spin of $\frac{7}{2}$ which results in eight hyperfine lines, each having zero field splittings between $0.05$ and $0.3$~T as predicted by the spin Hamiltonian. The spectroscopy identified other six hyperfine lines above 0.3~T with decreasing intensity. Since the intensity of these hyperfine interactions decreases with the line number (increasing magnetic field), the whole total number of lines could not be limited by 6. All the lines within two groups and between the groups themselves are almost parallel to each other as depicted in Fig.~\ref{interactionstrongfields2_vanadiumspinspin2}. Group (B) does not originates from any point on the positive frequency axis suggesting that it is not directly originated from the ZFS effect, i.e. ion-lattice coupling. This figure demonstrates that these two groups of hyperfine transitions are separated by approximately the ZFS of two types of Fe$^{3+}$ transitions $\Ket{\pm\frac{1}{2}}\rightarrow\Ket{+\frac{3}{2}}$ and $\Ket{-\frac{3}{2}}\rightarrow\Ket{\pm\frac{1}{2}}$. Thus, it is possible to hypothesise that the separation of Vanadium transitions into two groups of lines originates in coupling of the V$^{2+}$ single ions to a bath of Fe$^{3+}$ ions, i.e. spin-spin coupling. The spin-bath environments of other central spins (typically SQUIDs) could dominate over the oscillator bath environment at low temperatures\cite{Prokopev}. The probable mechanism of interaction involves one V$^{2+}$ ion nucleus and the iron spin bath which is the most abandoned impurity. 
 


\section*{Conclusion}

The proposed spectroscopy approach is demonstrated to be an ultra-sensitive tool in studying dilute impurities at low temperatures using high $Q$ WG modes in cylindrical single crystal resonators. The physical properties of ion impurities in high purity sapphire have been determined at 115 mK.
Using this approach it is possible to detect naturally occurring impurity ions in the purest single crystal dielectric resonators that can be in principle used for various high-$Q$ microwave applications. Spectroscopy revealed numerous single and two photon transitions of impurity ions, in particular Fe$^{3+}$, Cr$^{3+}$, and V$^{2+}$. The work demonstrates the possibility of QED in crystal cavities with natural impurities. Although the strong interaction regime has not been observed yet, due to the broadening of the Fe$^{3+}$ ESR. Broadening arises from strong spin-lattice coupling. The problem could be resolved by increasing the number of active ions by further annealing the crystal, that enhances the effective photon-spin coupling.


%

\end{document}